\documentclass[prl,reprint,twocolumn,showpacs,superscriptaddress]{revtex4-1}

\usepackage{bm,mathrsfs}
\usepackage{graphicx}
\usepackage{epsfig}
\usepackage{amsmath,bbm}
\usepackage{amsfonts,amssymb}
\usepackage{times}
\usepackage{dsfont}
\usepackage{enumitem}  
\usepackage{comment}
\usepackage[colorlinks,linkcolor=blue,citecolor=blue]{hyperref}
\newcommand{\eqn}[1]{
\begin{eqnarray}
	#1
\end{eqnarray}
}

\begin{document}

\title{Thermalization and Heating Dynamics in Open Generic Many-Body Systems}

\author{Yuto Ashida}
\affiliation{Department of Physics, University of Tokyo, 7-3-1 Hongo, Bunkyo-ku, Tokyo
113-0033, Japan}
\author{Keiji Saito}
\affiliation{Department of Physics, Keio university, Hiyoshi 3-14-1, Kohoku-ku, Yokohama, Japan}
\author{Masahito Ueda}
\affiliation{Department of Physics, University of Tokyo, 7-3-1 Hongo, Bunkyo-ku, Tokyo
113-0033, Japan}
\affiliation{RIKEN Center for Emergent Matter Science (CEMS), Wako, Saitama 351-0198, Japan
}

\begin{abstract}
The last decade has witnessed the remarkable progress in our understanding of thermalization in isolated quantum systems. Combining the eigenstate thermalization hypothesis with quantum measurement theory, we extend the framework of quantum thermalization to open many-body systems. 
A generic many-body system subject to continuous observation is shown to thermalize at a single trajectory level. We show that the nonunitary nature of quantum measurement causes several unique thermalization mechanisms that are unseen in isolated systems.
We present numerical evidence for our findings by applying our theory to specific models that can be experimentally realized in atom-cavity systems and  with quantum gas microscopy. Our theory provides a general  method to  determine an effective temperature of quantum many-body systems subject to the Lindblad master equation and thus should be  applicable to noisy dynamics or dissipative systems coupled to nonthermal Markovian environments as well as continuously monitored systems. Our work provides yet another insight into why thermodynamics emerges so universally. 
\end{abstract}
\maketitle

Statistical mechanics offers a universal framework to describe thermodynamic properties of a system involving many degrees of freedom \cite{29NV,JET57,MVB77,PA84,JRV85,SM94,DJM91,HT98}. Systems described by statistical mechanics can be divided into three distinct classes: (i) systems in contact with large thermal baths, (ii) isolated systems, and (iii) systems coupled to  nonthermal environments. Thermalization in the first class can be described by a phenomenological master equation in which  the detailed balance condition ensures that the system always relaxes to the Gibbs ensemble with the temperature of the thermal bath \cite{DEB74,SH78,RA79,AOC83,UW99,BHP02}.  The last decade has witnessed a considerable progress in our understanding of  thermalization in the second class  \cite{SP06,GS06,KC07,MSR07,RP08,CMF08,MCB11,NC12,AJS12,CF16,TM18,ADA15,TK16,RM08,RM09,BG10,SR13,SR14,KH14,BW14,KA15,LDA16,GJR18,YT18}, as promoted by quantum gas experiments \cite{TK06,ST12,GM12,KAM16,TY18}. 
In particular, the eigenstate thermalization hypothesis (ETH) \cite{SM94,DJM91,RM08,RM09,BG10,SR13,SR14,KH14,BW14,KA15,LDA16,GJR18,YT18} has emerged as a generic mechanism of thermalization under unitary dynamics of isolated quantum systems. The ETH has been numerically verified for a number of many-body Hamiltonians \cite{RM08,RM09,BG10,SR13,SR14,KH14,BW14,KA15,LDA16,GJR18,YT18} with notable exceptions of integrable \cite{RM07,CP11,FM13,SS14,FHLE16,LV16,FL17} or many-body localized systems \cite{RN15,RV16}.  

In class (iii), a coupling to a nonthermal environment violates the detailed balanced condition, as it permits arbitrary nonunitary processes such as continuous measurements \cite{PH10,GJP10,LTE14e,LTE14,LMD14,MKP14,YA15,ETJ15,WAC15,DS15A,YA16,DS16,MG16,YA17,SM17,YA17nc,KK17,MM18,LB18,JJS18} and engineered dissipation \cite{BA00,BK08,SD08,FV09,WY12,PD12,PD13,JC16,YA18,ZG18,BG13,PYS15,GT15,LR16,RB16,LHP17,TT17}. There, the bath temperature does not exist in general, and a number of fundamental questions arise.  
Does the system still thermalize and, if yes, in what sense? How are  steady states under such situations related to the thermal equilibrium of the system Hamiltonian? 
These questions are directly relevant to recent experiments realizing various types of controlled dissipations and measurements \cite{BG13,PYS15,GT15,LR16,RB16,LHP17,TT17} and to the foundations of open-system nonequilibrium statistical mechanics. 
The related questions were previously addressed in numerical studies of specific examples \cite{PH10,BZ14,SJ14,AD15,MG16}. Yet, model-independent, general understanding of thermalization and dynamics in open many-body systems is still elusive.

The aim of this Letter is to extend the framework of quantum thermalization to many-body systems coupled to Markovian environments permitted by quantum measurements and controlled dissipations. We consider open-system dynamics under continuous measurement process, i.e., weak and frequent repeated measurement, which can be realized by experimental setups of, e.g., atom-cavity systems \cite{BK10,WM12,SD14} and quantum gas microscopy \cite{BWS09,SJF10}.  Combining the ETH and quantum measurement theory, we derive a matrix-vector product expression of the time-averaged density matrix, and show that a generic many-body system under continuous observation will thermalize at a single trajectory level. The obtained density matrix can also be used to determine an effective temperature of open many-body systems governed by the Lindblad master equation. Our results  can thus be applied to dissipative many-body dynamics of a system coupled to a (not necessarily thermal) Markovian environment \cite{BK08,SD08,FV09,WY12,PD12,PD13,JC16,YA18,ZG18,BG13,PYS15,GT15,LR16,RB16,LHP17,TT17,BZ14} or under noisy unitary operations \cite{MJ12,PH13,CA17,BL17,HP16,NA17,NA18,vK18,CS18,MK18}.   
We also present numerical evidence of these findings by applying our theory to specific models. Our results give yet another insight into why thermodynamics emerges so universally.

\paragraph{Quantum many-body dynamics under measurement.\,---} 
We consider a generic (nonnintegrable) quantum many-body system subject to continuous observation. We assume that the system is initially prepared in a thermal equilibrium state $\hat{\rho}_{\rm eq}$, which is characterized by the mean energy $E_0$, or equivalently by the corresponding temperature $T_0=1/\beta_0$. We set $\hbar\!=\!1$ and $k_{\rm B}\!=\!1$ throughout this Letter. Following the standard theory of quantum measurement \cite{HC93}, we model a  measurement process as repeated indirect measurements. We start from a separable state  
\eqn{
\hat{\rho}_{{\rm tot}}(0) & = & \hat{\rho}_{{\rm eq}}\otimes\hat{P}_{0},
}
where $\hat{P}_{0}$ is a projection operator on the reference state of the meter. The system interacts with a meter during a time interval $\delta t$ via the total Hamiltonian $\hat{H}_{{\rm tot}}\!=\!\hat{H}+\hat{V}$,
where $\hat H$ is the many-body Hamiltonian of the measured system and $\hat{V}\!=\!v\sum_{m=1}^{M}\hat{L}_{m}\otimes\hat{A}_{m}$ describes the system-meter interaction. We assume that $\hat{L}_m$ is either a single local operator or the sum of local operators on the system and conserves the total particle number. We also assume that $\hat{A}_m$ acts on the state of the meter such that $\hat{P}_{l}\hat{A}_{m}=\delta_{lm}\hat{A}_{m}$, where $\hat{P}_{l}$'s are projection operators satisfying $\sum_{l=0}^{M}\hat{P}_{l}=1$.
After each interaction, we perform a projection measurement $\{\hat{P}_{l}\}$ on the meter to read out an outcome $l\!=\!0,1,\ldots M$. The meter is then reset to the reference state $\hat{P}_0$, which ensures that the dynamics is Markovian (see the Supplementary Material~\cite{SM1} for discussions on a non-Markovian case). For each measurement process, the meter exhibits either (i) a change in the state of the meter corresponding to outcome $m\!=\!1,2,\ldots M$ or (ii)  no change. The case (i) is referred to as a quantum jump process and accompanied by the following nonunitary mapping  
\eqn{\label{cptp}
{\cal E}_{m}(\hat{\rho})\!=\!{\rm Tr}_{\rm M}[\hat{P}_{m}\hat{U}(\delta t)\hat{\rho}_{{\rm tot}}\hat{U}^{\dagger}(\delta t)\hat{P}_{m}]
\!\simeq\!\gamma\delta t\hat{L}_{m}\hat{\rho}\hat{L}_{m}^{\dagger},
}
where  ${\rm Tr}_{\rm M}[\cdot]$ denotes the trace over the meter, $\hat{U}(\tau)\!=\!e^{-i\hat{H}_{\rm tot}\tau}$, and $\gamma\!=\!v^{2}\delta t{\rm Tr}_{A}[\hat{P}_0\hat{A}_{m}^{\dagger}\hat{A}_{m}]$ \footnote{For the sake of simplicity, we assume that the rate $\gamma$ is independent of an outcome $m$. A generalization to outcome-dependent rates is straightforward.}. In deriving the last expression in Eq.~\eqref{cptp}, we assume $\gamma\delta t\ll1$. The case (ii) is referred to as a no-count process, leading to
\eqn{
{\cal E}_{0}(\hat{\rho})\simeq(1-i\hat{H}_{{\rm eff}}\delta t)\hat{\rho}(1+i\hat{H}_{{\rm eff}}^{\dagger}\delta t),
}
where $\hat{H}_{{\rm eff}}\!=\!\hat{H}-i\hat{\Gamma}/2$  is an effective non-Hermitian Hamiltonian with $\hat{\Gamma}\!=\!\gamma\sum_{m}\hat{L}_{m}^{\dagger}\hat{L}_{m}$. Each outcome $l$ is obtained with a probability $p_{l}\!=\!{\rm Tr}[{\cal E}_{l}(\hat{\rho})]$. Taking the limit $\delta t\!\to\!0$ while keeping $v^2\delta t$ finite, the system exhibits a nonunitary stochastic evolution which is continuous in time and known as the quantum trajectory dynamics \cite{MU89,MU90,DR92,HC93,DJ92}. Each realization of a trajectory is characterized by a sequence of measurement outcomes and given as
\eqn{
\hat{\varrho}_{{\cal M}}(t;{\cal T})=\hat{\Pi}_{t;{\cal T}}^{{\cal M}}\hat{\rho}_{{\rm eq}}\hat{\Pi}_{t;{\cal T}}^{\dagger{\cal M}},
} 
where ${\cal M}\!=\!\left(m_{1},\ldots m_{n}\right)$ and ${\cal T}\!=\!\left(t_{1},\ldots t_{n}\right)$ denote the types and occurrence times of quantum jumps, and
$\hat{\Pi}_{t;{\cal T}}^{{\cal M}}\!=\!\prod_{i=1}^{n}[\hat{{\cal U}}(\Delta t_{i})\sqrt{\gamma}\hat{L}_{m_{i}}]\hat{{\cal U}}(t_{1})$ with $\Delta t_{i}\!=\!t_{i+1}-t_{i}$, $t_{n+1}\!=\!t$ and $\hat{{\cal U}}(\tau)\!=\!e^{-i\hat{H}_{{\rm eff}}\tau}$.

\paragraph{Statistical ensemble.\,---} 
We are interested in the thermalization process caused by the interplay between many-body dynamics and measurement backaction of continuous observation.
 We consider a situation in which the equilibration time in the measured many-body system is shorter than a typical waiting time between quantum jumps.  We ensure this by taking the limit $\gamma\!\to\!0$ while keeping $\gamma t$ finite. This guarantees that a finite number of quantum jumps typically occur during a given time interval $[0,t]$, but that the system still has not yet reached a steady state (such as an infinite-temperature state) even in the long-time regime.

For closed many-body systems, it has been argued that the equilibration time can be estimated as the Boltzmann time $\hbar/k_{\rm B}T$ with $T$ being the temperature of the system \cite{RP16,TRO18,WH17}. In the present context, these results imply the fast equilibration during no-jump process since the dominant contribution of $\hat{H}_{\rm eff}$ in the above limit is the (Hermitian) many-body Hamiltonian $\hat{H}$. Indeed, our numerical results presented below support this expectation, though we leave its mathematically rigorous proof open.  

When the waiting time exceeds the equilibration time, the exact state and its time-averaged density matrix are indistinguishable in terms of an expectation value of a physical observable. The reason is that the time-dependent elements of the density matrix make negligible contributions to an expectation value due to their rapid phase oscillations \cite{RP08}. This emergent time-independent feature indicates that the memory of the occurrence times $\cal T$ will be lost and expectation values of physical observables can be studied by the time-averaged density matrix
\eqn{
\hat{\varrho}_{{\cal M}}(t)=\int_{0}^{t}dt_{n}\cdots\int_{0}^{t_{2}}dt_{1}\hat{\varrho}_{{\cal M}}(t;{\cal T}).
}

\begin{figure}[b]
\includegraphics[width=80mm]{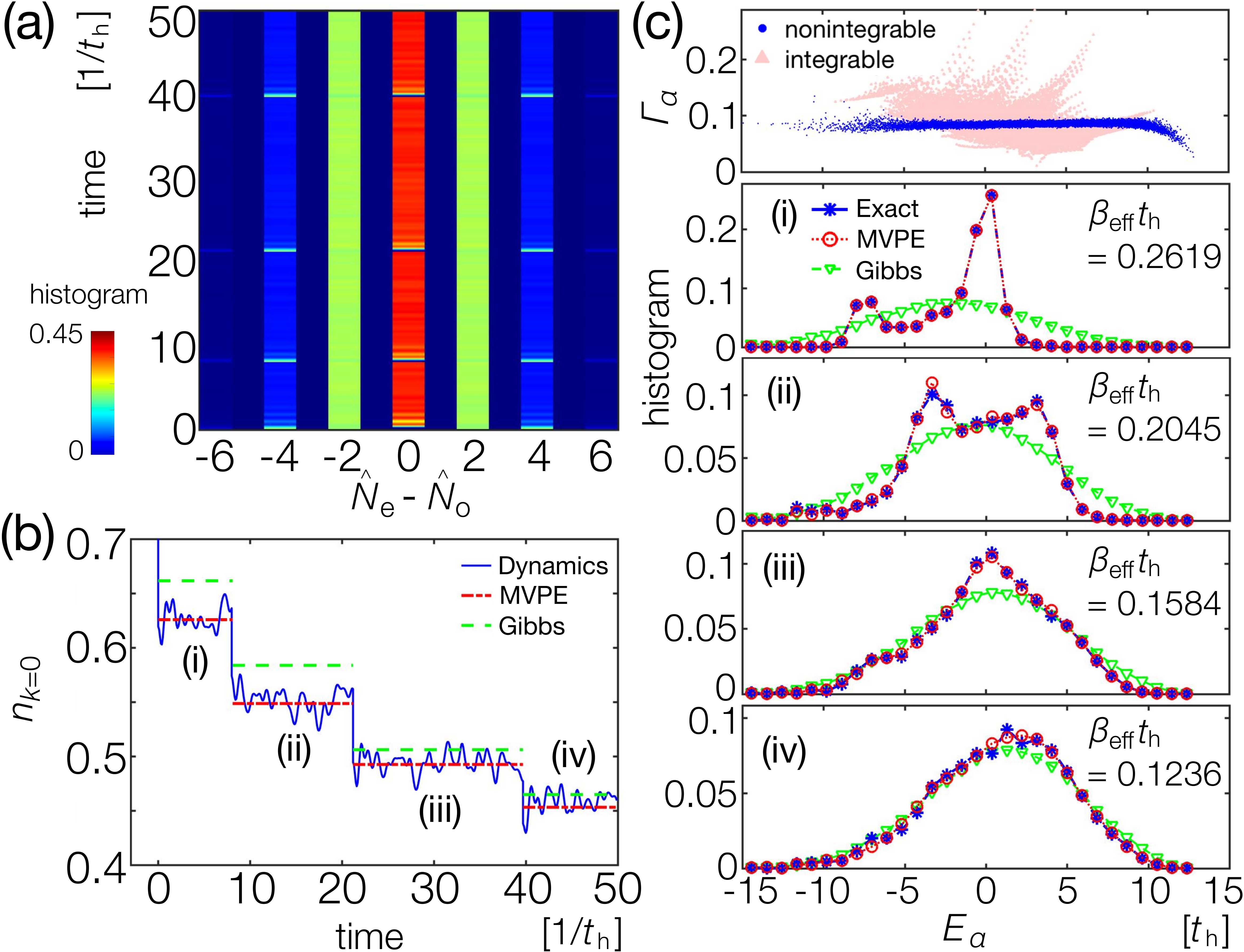} 
\caption{\label{fig1}
(a)  Time evolution of the distribution with the jump operator $\hat{L}\!=\!\sum_{l}(-1)^l\hat{n}_l$, which gives the difference $\hat{N}_{\rm e}-\hat{N}_{\rm o}$ of particle numbers at even and odd sites. Every time a quantum jump occurs, the distribution  peaks at $\hat{L}\!=\!\pm 4$ after which it rapidly relaxes toward an equilibrium profile due to the noncommutativity of $\hat{L}$ with the system Hamiltonian $\hat{H}$. (b) The corresponding dynamics of $\langle\hat{n}_{k=0}\rangle$. Superimposed is the prediction from the MVPE conditioned on a sequence of quantum jumps that have occurred by time $t$  (red dashed lines) and the Gibbs ensemble $\hat{\rho}_{{\beta}_{\rm eff}^{\cal M}}$ (green dashed lines).
(c) Diagonal values of the detection rate $\hat{\Gamma}$ (top panel) and the energy distributions after each jump (the other panels).
We set $\gamma=0.02$ and $t_{\rm h}\!=\!U\!=\!t_{\rm h}'\!=\!U'\!=\!1$ except for the integrable case in (c) where we use $t_{\rm h}\!=\!U\!=\!1$ and $t_{\rm h}'\!=\!U'\!=\!0$.
}
\end{figure}

To proceed with the calculation, we assume that  for each eigenstate $|E_a\rangle$ of $\hat{H}$ the expectation values of arbitrary few-body observables coincide with those of the corresponding Gibbs ensemble. This condition is generally believed to hold when the system satisfies the ETH \cite{LDA16}, as numerically supported for a number of Hamiltonians \cite{RM08,RM09,BG10,SR13,SR14,KH14,BW14,KA15,LDA16,GJR18,YT18}. The leading contribution in $\hat{\varrho}_{\cal M}$ can  be given upon the normalization as 
\eqn{\label{simple}
\hat{\rho}_{{\cal M}}=\frac{\Lambda_{{\cal M}}[\hat{\rho}_{{\rm eq}}]}{Z({\cal M})},
}
where we define $\Lambda_{{\cal M}}\!=\!\prod_{i=1}^{n}\left(\Lambda\circ{\cal L}_{m_{i}}\circ\Lambda\right)$ with ${\cal L}_{m}[\hat{O}]\!=\!\hat{L}_{m}\hat{O}\hat{L}_{m}^{\dagger}$, $\Lambda[\hat{O}]\!=\!\sum_{a}\hat{P}_{a}\hat{O}\hat{P}_{a}$ and $\hat{P}_a\!=\!|E_a\rangle\langle E_a|$, and $Z(\cal M)$ is a normalization constant.  While the non-Hermiticity in $\hat{H}_{\rm eff}$ can slightly modify the energy distribution, its contribution can be neglected in the thermodynamic limit \cite{SM1}. This follows from  strong suppression of fluctuations in the decay rate $\hat{\Gamma}$ among eigenstates that are close in energy (see, e.g., the top panel in Fig.~\ref{fig1}c). This suppression can be understood from the ETH \cite{SM94,LDA16}, which predicts the exponential decay of the fluctuations of the matrix elements in the energy basis in the thermodynamic limit. 

In the matrix representation, the ensemble~\eqref{simple} has a simple factorized form: 
\eqn{\label{matsimple}
\hat{\rho}_{{\cal M}}\propto\sum_{a}\left[{\cal V}_{m_{n}}\cdots{\cal V}_{m_{1}}p_{{\rm eq}}\right]_{a}\hat{P}_{a},
}
where we introduce the vector $(p_{\rm eq})_{a}\!=\!\langle E_a|\hat{\rho}_{\rm eq}|E_a\rangle$ and the matrix $({\cal V}_{m})_{ab}\!=\!|\langle E_{a}|\hat{L}_{m}|E_{b}\rangle|^{2}$. It follows from the cluster decomposition property \cite{SS14,SH15,KM14} of  thermal eigenstates for local operators $\hat{O}_{x,y}$ that 
\eqn{
\lim_{|x-y|\to\infty}{\rm Tr}[\hat{O}_{x}\hat{O}_{y}\hat{P}_{a}]-{\rm Tr}[\hat{O}_{x}\hat{P}_{a}]{\rm Tr}[\hat{O}_{y}\hat{P}_{a}]=0.
}
Then we can show \cite{SM1} that the standard deviation of energy  in the ensemble~\eqref{matsimple} is subextensive and thus its energy distribution is strongly peaked around the mean value $\overline{E}_{\cal M}$. We introduce an effective temperature $\beta_{\rm eff}^{\cal M}$ from the condition $\overline{E}_{\cal M}\!=\!{\rm Tr}[\hat{H}\hat{\rho}_{\beta_{\rm eff}^{\cal M}}]$ with $\hat{\rho}_{\beta}\!=\!e^{-\beta\hat{H}}/Z_\beta$ being the Gibbs ensemble. The ETH then guarantees that, if we focus on a few-body observable $\hat{O}$, $\hat{\rho}_{\cal M}$ is indistinguishable from the Gibbs ensemble:
\eqn{\label{gibbs}
{\rm Tr}[\hat{O}\hat{\rho}_{{\cal M}}]\simeq{\rm Tr}[\hat{O}\hat{\rho}_{\beta_{{\rm eff}}^{{\cal M}}}].
}
Here and henceforth we understand $\simeq$ to be the equality in the thermodynamic limit. Thus, a generic quantum system under a measurement process thermalizes by itself at  a single-trajectory level.

The derivation of the matrix-vector product ensemble (MVPE) in Eq.~\eqref{matsimple} is one of the main results in this Letter. In open many-body dynamics it is highly nontrivial to precisely estimate an effective temperature of the system. One usually has to design ad hoc techniques for each individual problem. In contrast, the MVPE provides a general and efficient way to determine an effective temperature under physically plausible assumptions as demonstrated later. If the ETH holds, any physical quantity can be calculated from the Gibbs ensemble at an extracted temperature. 

Before examining concrete examples, we discuss some general properties of thermalization under quantum measurement in comparison with thermalization in isolated systems.
 Firstly, since $\hat{H}$ has no local conserved quantities, it  satisfies $[\hat{H},\hat{L}_m]\!\neq\!0$ and thus the matrix ${\cal V}_m$ should change the energy distribution. It is this noncommutativity between the Hamiltonian and measurement operators that leads to heating or cooling  under measurement. 
Secondly, it is worthwhile to mention similarities and differences between Eq.~\eqref{matsimple} and the density matrix of isolated systems under slow time-dependent operations \cite{LDA16} or a sudden quench \cite{RM08,SL11,AP11,TNI15}. Both density matrices are diagonal in the energy basis and coefficients are represented in the matrix-vector product form. The unitarity inevitably leads to the doubly stochastic condition  of the transition matrix $\sum_{a}({\cal V})_{ab}=\sum_{b}({\cal V})_{ab}=1$, which causes the energy of the system to increase or stay constant \cite{ADA15,TK16,WT02}.
However, in the nonunitary evolution considered here, $\cal V$ cannot be interpreted as the transition matrix and the doubly stochastic condition is generally violated. This is why it is possible to cool down the system if one uses artificial (typically non-Hermitian) measurement operators $\hat{L}_{m}$ \cite{BK08,SD08}.

\paragraph{Numerical simulations.\,---}
To demonstrate our general approach, we consider a Hamiltonian $\hat{H}\!=\!\hat{K}+\hat{U}$ of hard-core bosons on an open one-dimensional lattice with nearest- and next-nearest-neighbor hopping and an interaction: $\hat{K}\!=\!-\sum_{l}(t_{{\rm h}}\hat{b}_{l}^{\dagger}\hat{b}_{l+1}\!+\!t'_{{\rm h}}\hat{b}_{l}^{\dagger}\hat{b}_{l+2}\!+\!{\rm H.c.})$ and $\hat{U}\!=\!\sum_{l}(U\hat{n}_{l}\hat{n}_{l+1}\!+\!U'\hat{n}_{l}\hat{n}_{l+2})$,
where $\hat{b}_l$ ($\hat{b}_{l}^\dagger$) is the annihiliation (creation) operator of a hard-core boson on site $l$ and $\hat{n}_{l}\!=\!\hat{b}_{l}^\dagger\hat{b}_{l}$. This model is, in general, nonintegrable and has been numerically confirmed to satisfy the ETH \cite{RM09,KH14,LDA16,YT18}. We set the system size and the total number of particles as $L_s\!=\!18$ and $N\!=\!6$. As a measurement process, we consider  $\hat{L}\!=\!\sum_{l}(-1)^{l}\hat{n}_{l}$ which can be implemented by monitoring photons leaking out of a cavity coupled to a certain collective mode of atoms \cite{MG16}.

To test the validity of the MVPE \eqref{matsimple} for describing open many-body dynamics, it suffices to use an energy eigenstate as the initial state. Results for a general initial distribution $p_{\rm eq}$ can be obtained as merely a linear sum of the results for individual eigenstates.
To be specific, we start from an eigenstate $|E_{0}\rangle$ corresponding to the initial temperature $T_0\!=\!3t_{\rm h}$. Without loss of generality, we choose the first detection time of a quantum jump as $t=0$.

Figure~\ref{fig1} shows a typical realization of quantum dynamics under the measurement. Figure~\ref{fig1}a plots the time evolution of the distribution of $\hat{L}\!=\!\sum_{l}(-1)^l\hat{n}_{l}$, which is the difference of particle numbers at even and odd sites. Each detection creates a cat-like post-measurement state having large weights on $\hat{L}\!=\!\pm 4$. It then quickly decays into a thermal state since $\hat{H}$ does not commute with $\hat{L}$. In Fig.~\ref{fig1}b, the corresponding dynamics of $\langle\hat{n}_{k=0}\rangle$ is compared with the predictions from the MVPE $\hat{\rho}_{\cal M}$ and the Gibbs ensemble $\hat{\rho}_{\beta_{\rm eff}^{\cal M}}$.
We find an excellent  agreement between the exact values and the MVPE.
Note that the MVPE is time independent by definition; the plotted values correspond to the MVPE conditioned on a sequence of quantum jumps that have occurred by time $t$. 
 Figure~\ref{fig1}c shows the diagonal values ${\Gamma}_{a}$ of the detection rate (top panel) and energy distributions after each jump (the other panels). The latter shows a rapid collapse of the energy distribution into that of the Gibbs ensemble after only a few jumps. The similar results are also found in numerical simulations for a local density measurement $\hat{L}_{i}=\hat{n}_{i}$ \cite{SM1}, which is directly relevant to quantum gas microscopy.

\begin{figure}[t]
\includegraphics[width=80mm]{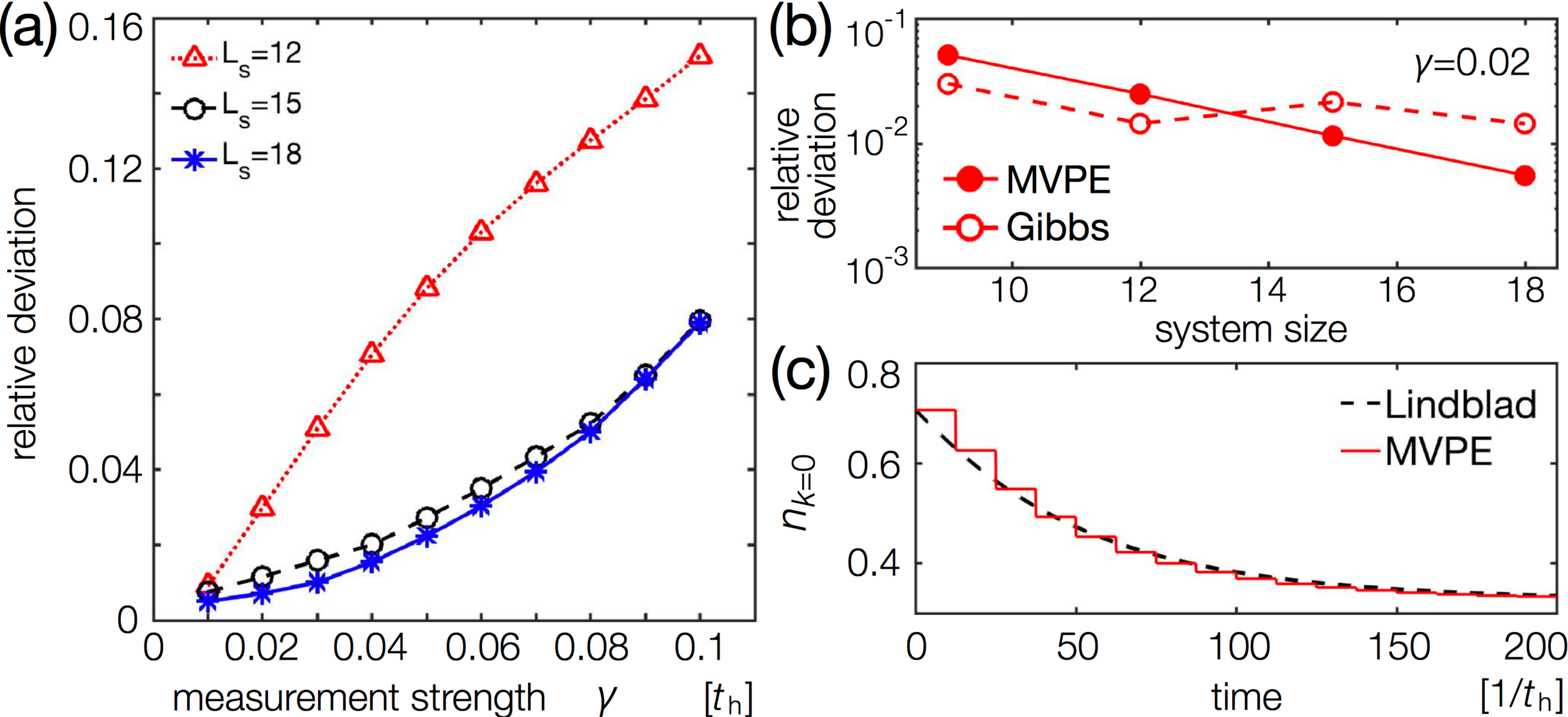} 
\caption{\label{fig2}
(a) Relative deviations of  MVPE predictions from time-averaged values of $\hat{n}_{k=0}$  plotted against measurement strength $\gamma$ for different system sizes $L_s$. (b) Finite-size scaling analyses of the relative deviations of  MVPE (filled circles) and the corresponding Gibbs ensemble (open circles) from  time-averaged values of $\langle\hat{n}_{k=0}\rangle$. (c) Comparison between the MVPE predictions (red solid lines) and the Lindblad dynamics (black dashed curve) for $\hat{n}_{k=0}$ with $L_s=18$. 
We use $t_{\rm h}\!=\!U\!=\!t_{\rm h}'\!=\!U'\!=\!1$ and set  $\gamma=0.02$ in (b) and  (c).
}
\end{figure}

Figure~\ref{fig2} shows relative deviations of the MVPE predictions from time-averaged values of $\hat{n}_{k=0}$ with varying  system-meter coupling $\gamma$. Finite-size scaling analyses indicate that the relative deviations become exponentially small with increasing the system size for small $\gamma$ (see e.g., Fig.~\ref{fig2}b), while they no longer converge for  larger values (typically $\gamma\gtrsim 0.08t_{\rm h}$) in which the minimally destructive limit breaks down.
A relatively slow convergence of the  Gibbs ensemble predictions in Fig.~\ref{fig2}b can be attributed to  broad energy distributions and large fluctuations in diagonal elements of $\hat{n}_{k=0}$ in finite-size systems (see Ref.~\cite{SJ14} for a similar observation).

\paragraph{Application to many-body Lindblad dynamics.\,---}
Having established the validity of the MVPE, we now discuss its application to the Lindblad dynamics. The quantum trajectory dynamics offers a numerical method to solve the Lindblad master equation \cite{BHP02,AD15}:
\eqn{\label{lindblad}
\frac{d\hat{\rho}}{dt}={\cal L}[\hat{\rho}]=-i(\hat{H}_{{\rm eff}}\hat{\rho}-\hat{\rho}\hat{H}_{{\rm eff}}^{\dagger})+\gamma\sum_{m}\hat{L}_{m}\hat{\rho}\hat{L}_{m}^{\dagger},
}
where $\hat{\rho}(t)\!=\!\sum_{{\cal M}}\hat{\varrho}_{{\cal M}}(t)$ is the density matrix averaged over all trajectories.
Equation~\eqref{lindblad} can describe the temporal evolution of a system weakly coupled to its environment \cite{BHP02} or  a system under noisy unitary operations \cite{MJ12,PH13,CA17,BL17,MK18}.
 Yet, especially for a many-body system, it is often very demanding to take the ensemble average due to a vast number of possible trajectories.  

For the case of a translationally invariant Hamiltonian $\hat{H}$ and a local operator $\hat{L}_m$, our approach suggests a simple way to overcome the above difficulty. In this case, the matrix ${\cal V}_m$ is independent of a spatial label $m$ and thus the MVPE in Eq.~\eqref{matsimple} is characterized by the number $n$ of quantum jumps alone: $\hat{\rho}_{n}\!\propto\!\sum_{a}\!\left[{\cal V}^{n}p_{{\rm eq}}\right]_{a}\!\hat{P}_{a}$. As the detection rate $\hat{\Gamma}$ of quantum jumps consists of few-body observables,  the distribution of $n$ is sharply peaked around the mean value $\overline{n}$ if the ETH holds. These observations lead to
\eqn{\label{lindbladsimple}
{\rm Tr}[\hat{O}e^{{\cal L}t}\hat{\rho}_{{\rm eq}}]\simeq{\rm Tr}[\hat{O}\hat{\rho}_{\overline{n}_t}]\simeq{\rm Tr}[\hat{O}\hat{\rho}_{\beta_{{\rm eff}}^{\overline{n}_t}}],
}
where $\overline{n}_t$ is the mean number of quantum jumps during $[0,t]$ that can be determined from the implicit relation $t\!\simeq\!\sum_{n=0}^{\overline{n}_t}1/\Gamma_{n}$ with $\Gamma_{n}\!=\!{\rm Tr}[\hat{\Gamma}\hat{\rho}_{n}]$, and $\beta_{\rm eff}^{\overline{n}_t}$ is the corresponding effective temperature. Thus, expectation values of physical observables in the many-body Lindblad dynamics agree with those predicted from the typical MVPE or the Gibbs ensemble at an appropriate effective temperature. 
Since solving Eq.~\eqref{lindblad} requires the diagonalization of a $D^2\!\times\!D^2$ Liouvillean with $D$ being the dimension of the Hilbert space, our approach~\eqref{lindbladsimple}  allows a significant simplification of the problem. We have applied our approach to the Lindblad dynamics of the above lattice model and demonstrated the relation~(\ref{lindbladsimple}) aside from stepwise finite-size corrections (Fig.~\ref{fig2}c).

\paragraph{Summary and Discussions.\,---}
Combining the ideas of the ETH and quantum measurement theory,
we find that a generic quantum many-body system under continuous observation thermalizes at a single trajectory level. We have presented the matrix-vector product ensemble~\eqref{matsimple}, which can quantitatively describe the dynamics and give an effective temperature of an open quantum many-body system. This  can also be used to analyze a many-body Lindblad master equation and thus should have broad applicability to dissipative \cite{BK08,SD08,FV09,WY12,PD12,PD13,JC16,YA18,ZG18,BG13,PYS15,GT15,LR16,RB16,LHP17,TT17} or noisy Markovian systems \cite{MJ12,PH13,CA17,BL17,MK18}, in addition to continuously monitored ones. These findings are supported by numerical simulations of nonintegrable systems under continuous measurement, which can be experimentally realized in atom-cavity systems and by quantum gas microscopy.  

The present study opens several  research directions.
Firstly, it is intriguing to elucidate thermalization at the trajectory level when the system Hamiltonian is integrable \cite{RM07,CP11,FM13,SS14,FHLE16,LV16,FL17}. 
Under measurements, quantum jumps act as weak integrability-breaking perturbations and, if their effects are insignificant, we expect prethermalization \cite{RM07,TM18}, i.e., a phenomenon in which  observables approach  quasistationary values consistent with the generalized Gibbs ensemble \cite{RM07}. Ultimate thermalization will happen when quantum jumps sufficiently mix the distribution, leading to the unbiased probability weights on nonthermal rare states admitted in the weak variant of the ETH \cite{BG10,TM16,IE17,TM18}. We present our first attempt to outline this scenario in the Supplementary Material~\cite{SM1} and leave a detailed analysis as an interesting future problem. Another important system is a many-body localized system \cite{RN15,RV16} where even the weak ETH can be violated \cite{IJZ16}. Secondly, it remains an important problem to extend our analysis to  non-Markovian open dynamics \cite{BHP16}. While an application of a non-Markovian trajectory approach \cite{PJ08} to a many-body system is challenging in general, our MVPE approach may  still be useful if the support of a jump operator is restricted  \cite{SM1}.
Thirdly, it is interesting to explore possible connections between the predictions made in the random unitary circuit dynamics \cite{HP16,BL17,NA17,NA18,vK18,CS18} and the nonintegrable open trajectory dynamics studied here. They share several intriguing similarities; they satisfy the locality,  have no energy conservation, thus relaxing to the infinite-temperature state, and obey the Lindblad master equation upon the ensemble average (at least) in a certain case \cite{BL17}. It is particularly interesting to test the predicted scrambling dynamics \cite{NA18,vK18} or the Kardar-Parisi-Zhang universal behavior \cite{NA17} in the present setup.

We are grateful to Zala Lenarcic, Igor Mekhov, Takashi Mori and Hannes Pichler for useful discussions. We also thank Ryusuke Hamazaki for valuable suggestions, especially on the cluster decomposition property. We acknowledge support from KAKENHI Grant No.~JP18H01145 and a Grant-in-Aid for Scientific Research on Innovative Areas ``Topological Materials Science" (KAKENHI Grant No.~JP15H05855) from the Japan Society for the Promotion of Science (JSPS), and ImPACT Program of Council for Science, Technology and Innovation (Cabinet Office, Government of Japan). Y.~A. acknowledges support from JSPS (Grant No.~JP16J03613). 

\bibliography{reference}

\widetext
\pagebreak
\begin{center}
\textbf{\large Supplementary Materials}
\end{center}

\renewcommand{\theequation}{S\arabic{equation}}
\renewcommand{\thefigure}{S\arabic{figure}}
\renewcommand{\bibnumfmt}[1]{[S#1]}
\setcounter{equation}{0}
\setcounter{figure}{0}

\subsection{Derivation of the matrix-vector product form of the time-averaged density matrix}
We here provide technical details on the derivation of Eq.~(7) in the main text which is the matrix-product form of the density matrix in open many-body dynamics. Suppose that a many-body system is perturbed by a quantum jump. Then, after an equilibration time (which is typically of the order of the Boltzmann time \cite{RP16,TRO18,WH17}), one can no longer  distinguish the exact time-evolved state from its time-averaged  density matrix in terms of an expectation value of a physical observable \cite{RP08}. This is because that the fast-oscillating, time-dependent terms in the density matrix cancel out and make only subleading contributions to the expectation value. In our consideration, this emergent time-independent feature indicates the loss of memory of quantum-jump times, allowing one to study values of physical observables by
the following time-averaged density matrix  (see Eq.~(5) in the main text)
\eqn{\label{sm1}
\hat{\varrho}_{{\cal M}}(t)=\int_{0}^{t}dt_{n}\cdots\int_{0}^{t_{2}}dt_{1}\prod_{i=1}^{n}[\hat{{\cal U}}(\Delta t_{i})\sqrt{\gamma}\hat{L}_{m_{i}}]\hat{{\cal U}}(t_{1})\hat{\rho}_{{\rm eq}}\hat{{\cal U}}^{\dagger}(t_{1})\prod_{i=1}^{n}[\sqrt{\gamma}\hat{L}_{m_{i}}^{\dagger}\hat{{\cal U}}^{\dagger}(\Delta t_{i})],
}
where the time average is taken over all possible occurrence times of quantum jump events. To rewrite this, we expand a non-Hermitian effective Hamiltonian $\hat{H}_{\rm eff}$ as
$\hat{H}_{{\rm eff}}=\sum_{a}\Lambda_{a}|\Lambda_{a}^{{\rm R}}\rangle\langle\Lambda_{a}^{{\rm L}}|$, where $\Lambda_{a}$ is a complex eigenvalue, and the right (left) eigenstates $|\Lambda_a^{\rm R}\rangle$ ($|\Lambda_a^{\rm L}\rangle$)  satisfy the orthonormal condition
$\langle\Lambda_{a}^{{\rm R}}|\Lambda_{b}^{{\rm L}}\rangle=\delta_{ab}.$
We then insert the relation 
$\hat{{\cal U}}(\tau)=e^{-i\hat{H}_{\rm eff}\tau}=\sum_{a}e^{-i\Lambda_{a}\tau}|\Lambda_{a}^{{\rm R}}\rangle\langle\Lambda_{a}^{{\rm L}}|$ into Eq.~(\ref{sm1}), obtaining
\eqn{\label{sm2}
\hat{\varrho}_{{\cal M}}(t)=\sum_{\{a_{i}\}\{b_{i}\}}{\cal F}(t;\{a_{i}\},\{b_{i}\})\prod_{i=1}^{n}[\sqrt{\gamma}(V_{m_{i}})_{a_{i+1}a_{i}}](P_{{\rm eq}})_{a_{1}b_{1}}\prod_{i=1}^{n}[\sqrt{\gamma}(V_{m_{i}}^{\dagger})_{b_{i}b_{i+1}}]|\Lambda_{a_{n+1}}^{{\rm R}}\rangle\langle\Lambda_{b_{n+1}}^{{\rm L}}|,
}
where we introduce the matrices $V_m$ and $P_{\rm eq}$ as
\eqn{
(V_{m})_{ab}=\langle\Lambda_{a}^{{\rm L}}|\hat{L}_{m}|\Lambda_{b}^{{\rm R}}\rangle,\;\;\;(P_{{\rm eq}})_{ab}=\langle\Lambda_{a}^{{\rm L}}|\hat{\rho}_{{\rm eq}}|\Lambda_{b}^{{\rm L}}\rangle,
}
and $\cal F$ involves the time integration of the exponential factor 
\eqn{
{\cal F}(t;\{a_{i}\},\{b_{i}\})=e^{-i(\Lambda_{a_{n+1}}-\Lambda_{b_{n+1}}^{*})t}\int_{0}^{t}dt_{n}\cdots\int_{0}^{t_{2}}dt_{1}e^{-i\sum_{i=1}^{n}\Delta_{i}t_{i}}
}
with
\eqn{
\Delta_{i}=\Lambda_{a_{i}}-\Lambda_{b_{i}}^{*}-(\Lambda_{a_{i+1}}-\Lambda_{b_{i+1}}^{*}).
}
To extract generic features of the trajectory many-body dynamics described by Eq.~(\ref{sm2}), we consider the limit of minimally destructive observation, i.e., we take $\gamma\to 0$ while  keeping $\gamma t=\mu$ finite. Here, $\mu$ characterizes the mean number of jumps during a given time interval $[0,t]$. We also assume that, for each many-body eigenstate $|E_a\rangle$ of the system Hamiltonian $\hat{H}$, the expectation value of any few-body observable coincides with that over the corresponding Gibbs ensemble. This assumption is known as the eigenstate thermalization hypothesis (ETH) and  has been numerically verified for a number of systems \cite{RM08,RM09,BG10,SR13,SR14,KH14,BW14,KA15,LDA16,GJR18,YT18}. 

Assuming the minimally destructive limit and the ETH, we can achieve several simplifications. Firstly, we have only to take into account the leading contributions of the integral $\cal F$ in the limit of $t\to \infty$, i.e., the terms with $a_i=b_i$  ($i=1,2,\ldots,n+1$) for which ${\rm Re}[\Delta_{i}]=0$; the other terms are suppressed due to their rapid phase oscillations. We note that the condition ${\rm Re}[\Delta_{i}]=0$ uniquely leads to $a_i=b_i$ since we assume the equilibrium initial state, i.e., the density matrix diagonal in the energy basis, and the absence of degeneracy in the energy spectrum.  Secondly, the eigenstates $|\Lambda_{a}^{\rm R,L}\rangle$ are replaced by those $|E_a\rangle$ of the system Hamiltonian $\hat{H}$, as the minimally destructive limit ensures the vanishingly small non-Hermiticity $\gamma\to 0$ in the effective Hamiltonian. Accordingly, to express an imaginary part $\Gamma_a$ of an eigenvalue $\Lambda_a$, we can use the perturbative result $\Gamma_a=\gamma\langle E_a|\sum_m\hat{L}_m^{\dagger}\hat{L}_m|E_a\rangle$. These simplifications lead to
\eqn{\label{smtemp1}
\hat{\varrho}_{{\cal M}}(t)\simeq\sum_{\{a_{i}\}}e^{-\mu\tilde{\Gamma}_{a_{n+1}}}\int_{0}^{\mu}d\mu_{n}\cdots\int_{0}^{\mu_{2}}d\mu_{1}e^{-\sum_{i=1}^{n}\delta\tilde{\Gamma}_{a_{i}}\mu_{i}}({\cal V}_{m_{n}})_{a_{n+1}a_{n}}\cdots({\cal V}_{m_{1}})_{a_{2}a_{1}}(p_{{\rm eq}})_{a_{1}}|E_{a_{1}}\rangle\langle E_{a_{1}}|,
}
where ${\cal V}_m$ and $p_{\rm eq}$ are matrices whose elements are defined by
\eqn{
({\cal V}_{m})_{ab}=|\langle E_{a}|\hat{L}_{m}|E_{b}\rangle|^{2},\;\;\;(p_{{\rm eq}})_{a}=\langle E_{a}|\hat{\rho}_{{\rm eq}}|E_{a}\rangle,
}
and we introduce variables 
\eqn{
\mu_{i}&\equiv&\gamma t_{i},\\
\tilde{\Gamma}_{a}&\equiv&\Gamma_{a}/\gamma=\langle E_{a}|\sum_{m}\hat{L}_{m}^{\dagger}\hat{L}_{m}|E_{a}\rangle,\\
\delta\tilde{\Gamma}_{a_{i}}&\equiv&\tilde{\Gamma}_{a_{i}}-\tilde{\Gamma}_{a_{i+1}}.
}
To achieve further simplification, we note the fact that the off-diagonal elements $({\cal V}_{m})_{ab}$ vanish exponentially fast with increasing the energy difference $\omega=|E_a-E_b|$ \cite{AD15,TK16}. Thus, the dominant contributions to Eq.~(\ref{smtemp1}) are made from matrix-vector products for the elements $a_{i}$ and $a_{i+1}$ that are close in energy, i.e., the elements satisfying $|E_{a_{i}}-E_{a_{i+1}}|/|E_{a_{i}}+E_{a_{i+1}}|\ll 1$. For such elements,  the ETH guarantees that the fluctuation of the decay rate is strongly suppressed $|\Gamma_{a_i}-\Gamma_{a_{i+1}}|/|\Gamma_{a_i}+\Gamma_{a_{i+1}}|\ll 1$, as we consider physical jump operators $\hat{L}_m$ consisting of few-body operators. We thus neglect $\delta\tilde{\Gamma}_{a}$'s in Eq.~(\ref{smtemp1}), leading to
\eqn{\label{smtemp2}
\hat{\varrho}_{{\cal M}}(t)\simeq\frac{\mu^{n}}{n!}\sum_{a}e^{-\mu\tilde{\Gamma}_{a}}[{\cal V}_{m_{n}}\cdots{\cal V}_{m_{1}}p_{{\rm eq}}]_{a}|E_{a}\rangle\langle E_{a}|.
}
Finally, while successive multiplications of matrices ${\cal V}_m$ on the initial distribution $p_{\rm eq}$ can eventually change the mean energy $\overline{E}$ by an extensive amount, they still keep the energy fluctuation subextensive as shown in the next section. In other words, the energy distribution is strongly peaked around the mean value during each time interval between jump events. The ETH then guarantees that the distribution of the detection rate $\hat{\Gamma}$ is also strongly peaked and its fluctuation around the mean value  is vanishingly small in the thermodynamic limit. We thus  replace $\tilde{\Gamma}_{a}$ in Eq.~\eqref{smtemp2} by its mean value $\overline{\tilde{\Gamma}}$ in the final distribution $(\prod_{i=1}^{n}{\cal V}_{m_{i}})p_{\rm eq}$, and arrive at the following simple expression of the density matrix:
\eqn{
\hat{\varrho}_{{\cal M}}(t)\simeq\frac{\mu^{n}}{n!}e^{-\mu\overline{\tilde{\Gamma}}}\sum_{a}[{\cal V}_{m_{n}}\cdots{\cal V}_{m_{1}}p_{{\rm eq}}]_{a}|E_{a}\rangle\langle E_{a}|,
}
which gives Eqs.~(6) and (7) in the main text after normalization. 
Since the distribution in the energy basis is often rather broad, in finite-size systems, it is in practice useful to employ the expression~\eqref{smtemp2} as a matrix-vector product ensemble especially when the diagonal elements of the detection rate $\Gamma_a$ vary significantly as a function of energy.  

We remark about a possible extension of our MVPE approach to a non-Markovian case. When the non-Markovian dynamics is local in time, it can be described by the Lindblad-type  master equation but with time-dependent jump operators $\hat{L}_{m}(t)$ and coefficients $\gamma_{m}(t)$ \cite{BHP16}. Here the coefficient $\gamma_{m}(t)$ can be negative in a non-Markovian case, which makes it difficult to apply the standard quantum trajectory approach to the non-Markovian master equation. Nevertheless, it is possible to develop an analogue of quantum trajectory approach even in  a non-Markovian case \cite{PJ08} by introducing a ``backward" jump operator $\hat{D}_{\alpha\to\alpha'}^{m}$ for $\gamma_{m}(t)<0$. Here, $\alpha$ and $\alpha'$ represent a source state  and a target state, respectively, and the operator acts on the state as $|\psi_{\alpha'}(t+\delta t)\rangle=\hat{D}^{m}_{\alpha\to\alpha'}(t)|\psi_{\alpha}(t)\rangle$ such that $|\psi_\alpha (t)\rangle=\hat{L}_{m}(t)|\psi_{\alpha'}(t)\rangle/||\hat{L}_{m}(t)|\psi_{\alpha'}(t)\rangle||$. If one takes the minimally destructive limit $|\gamma_{m}(t)|\to 0$, we expect that our approximation of just taking into account the diagonal elements in the density matrix is still valid and thus the  MVPE can, in principle, be extended to the non-Markovian trajectory dynamics described above. Yet, for a negative coefficient $\gamma_m<0$, the number of matrices $({\cal D}_{\alpha\to\alpha'}^m)_{ab}\equiv|\langle E_a|\hat{D}_{\alpha\to\alpha'}^m|E_b\rangle|^2$ we must handle grows exponentially with respect to the system size due to exponentially large possibilities of the labels $\alpha$ and $\alpha'$. It is this difficulty that appears to make the MVPE approach impractical for a non-Markovian case. Nevertheless, we may still use it in a special case for which the support of a jump operator does not grow exponentially (owing to, e.g., a certain symmetry of a jump operator) and thus the total number of nonvanishing matrices ${\cal D}_{\alpha\to\alpha'}^m$ is still small and thus tractable.

\subsection{Subextensive energy fluctuation in the matrix-vector product ensemble}
In this section, we show that energy fluctuation in the following matrix-vector product ensemble is subextensive:
\eqn{\label{mvpe}
\hat{\rho}_{{\cal M}}=
\frac{\Lambda_{{\cal M}}[\hat{\rho}_{{\rm eq}}]}{Z({\cal M})}
=
\frac{1}{Z({\cal M})}\sum_{a}[{\cal V}_{m_{n}}\cdots{\cal V}_{m_{1}}p_{{\rm eq}}]_{a}|E_{a}\rangle\langle E_{a}|.
}
Here, we introduce $\Lambda_{{\cal M}}\!=\!\prod_{i=1}^{n}\left(\Lambda\circ{\cal L}_{m_{i}}\circ\Lambda\right)$ with ${\cal L}_{m}[\hat{O}]\!=\!\hat{L}_{m}\hat{O}\hat{L}_{m}^{\dagger}$, $\Lambda[\hat{O}]\!=\!\sum_{a}\hat{P}_{a}\hat{O}\hat{P}_{a}$ and $\hat{P}_a\!=\!|E_a\rangle\langle E_a|$, and $Z({\cal M})=\sum_{a}[{\cal V}_{m_{n}}\cdots{\cal V}_{m_{1}}p_{{\rm eq}}]_{a}$ is a normalization constant. 

We assume that thermal eigenstates satisfy the cluster decomposition property (CDP) (c.f. Eq.~(8) in the main text), which is the fundamental property that lies at the heart of quantum many-body theory \cite{SS14,SH15,KM14,FHLE16} and should hold true for any physical states with only a few exceptions such as (long-lived) macroscopic superposition states \cite{FHLE16}. From this assumption together with the ETH, it follows that any diagonal ensemble with a strongly peaked energy distribution satisfies the CDP in the thermodynamic limit. 
In particular, the initial thermal equilibrium state $\hat{\rho}_{\rm eq}$ also satisfies the CDP since its energy fluctuation (i.e., the standard deviation) is subextensive by definition.

Below we show that if  energy fluctuation in an ensemble $\hat{\rho}$ diagonal in the energy basis is subextensive and thus $\hat{\rho}$ satisfies the CDP, then a post-measurement ensemble $\hat{\rho}_{m}\propto\Lambda_{m}[\hat{\rho}]$ after a single quantum jump with $\Lambda_{m}=\Lambda\circ{\cal L}_{m}\circ\Lambda$ also satisfies these conditions.
It then follows from the induction that an energy fluctuation of the density matrix~\eqref{mvpe} is also subextensive.

\subsubsection*{Local measurement}
We first show the subextensiveness of  energy fluctuation in the post-measurement ensemble $\hat{\rho}_m$ for a local measurement, in which a measurement operator $\hat{L}_m$ acts on a local spatial region.
The variance of energy  is given as
\eqn{\label{ergf}
(\Delta E)^{2}={\rm Tr}[\hat{H}^{2}\hat{\rho}_{m}]-({\rm Tr}[\hat{H}\hat{\rho}_{m}])^{2}=\frac{1}{(Z(m))^{2}}\left[Z(m){\rm Tr}[\hat{H}^{2}{\cal L}_{m}[\hat{\rho}]]-\left({\rm Tr}[\hat{H}{\cal L}_{m}[\hat{\rho}]]\right)^{2}\right],
}
where $Z(m)={\rm Tr}[\Lambda_{m}[\hat{\rho}]]$ is a normalization constant.
We express the Hamiltonian and measurement operators as sums of local operators:
\eqn{
\hat{H}=\sum_{x}\hat{h}_{x},\;\;\;\hat{L}_{m}=\sum_{x\in{\cal D}_{m}}\hat{l}_{x},
}
where ${\cal D}_m$ denotes a local spatial region on which $\hat{L}_m$ acts.
To rewrite Eq.~\eqref{ergf}, we calculate the quantity
\eqn{\label{delxy}
\Delta_{xy} & \equiv&\frac{1}{(Z(m))^{2}}\left[Z(m){\rm Tr}[\hat{h}_{x}\hat{h}_{y}{\cal L}_{m}[\hat{\rho}]]-{\rm Tr}[\hat{h}_{x}{\cal L}_{m}[\hat{\rho}]]{\rm Tr}[\hat{h}_{y}{\cal L}_{m}[\hat{\rho}]]\right]\\
 & =&\frac{1}{(\langle\hat{L}_{m}^{\dagger}\hat{L}_{m}\rangle)^{2}}\left[\langle\hat{L}_{m}^{\dagger}\hat{L}_{m}\rangle\langle\hat{L}_{m}^{\dagger}\hat{h}_{x}\hat{h}_{y}\hat{L}_{m}\rangle-\langle\hat{L}_{m}^{\dagger}\hat{h}_{x}\hat{L}_{m}\rangle\langle\hat{L}_{m}^{\dagger}\hat{h}_{y}\hat{L}_{m}\rangle\right],
}
where we denote ${\rm Tr}[\cdot\hat{\rho}]\equiv\langle\cdot\rangle$. 
Using the condition $[\hat{l}_{x},\hat{h}_{y}]=0$ for $x\neq y$, we obtain  in the limit $|x-y|\to \infty$
\eqn{
\langle\hat{L}_{m}^{\dagger}\hat{h}_{x}\hat{h}_{y}\hat{L}_{m}\rangle=\langle\hat{L}_{m}^{\dagger}\hat{L}_{m}\hat{h}_{x}\hat{h}_{y}\rangle&\simeq&\langle\hat{L}_{m}^{\dagger}\hat{L}_{m}\rangle\langle\hat{h}_{x}\rangle\langle\hat{h}_{y}\rangle,\\
\langle\hat{L}_{m}^{\dagger}\hat{h}_{x}\hat{L}_{m}\rangle\langle\hat{L}_{m}^{\dagger}\hat{h}_{y}\hat{L}_{m}\rangle&\simeq&\langle\hat{L}_{m}^{\dagger}\hat{L}_{m}\rangle^{2}\langle\hat{h}_{x}\rangle\langle\hat{h}_{y}\rangle.
}
Here, we use the CDP of $\hat{\rho}$ in deriving the last expressions.
We thus obtain $\lim_{|x-y|\to\infty}\Delta_{xy}=0$.
It follows that the standard deviation of  energy is subextensive:
\eqn{
\lim_{V\to \infty}\frac{\Delta E}{V}=\lim_{V\to\infty}\frac{\sqrt{\sum_{xy}\Delta_{xy}}}{\sum_{xy}1}=0.
}
In particular, it is physically plausible to assume that $\Delta_{xy}$ decays exponentially fast or at least faster than $V^{-1}$ in the thermodynamic limit. Under this condition, we obtain the square-root scaling:
\eqn{
\Delta E\simeq\sqrt{\sum_{x}\Delta_{xx}}\propto {\cal O}(\sqrt{V}).
}
\subsubsection*{Global measurement}
We next consider a global measurement, in which a measurement operator acts on an entire region of the system. As ${\cal D}_m$ is independent of  label $m$, we abbreviate a label and denote a measurement operator as $\hat{L}=\sum_{z}\hat{l}_z$ for the sake of simplicity. It turns out that we need to discuss the two different cases separately depending on whether or not the expectation value $\lim_{V\to\infty}\langle\hat{L}\rangle/V$ vanishes in the thermodynamic limit.

We first consider the case in which $\langle\hat{L}\rangle$ scales as
\eqn{
\langle\hat{L}\rangle=\langle\sum_{z}\hat{l}_{z}\rangle\propto {\cal O}(V),
}
so that  $\lim_{V\to\infty}\langle\hat{L}\rangle/V$ does not vanish.
From the CDP of $\hat{\rho}$, the leading term in $\Delta_{xy}$ defined in Eq.~\eqref{delxy} can be estimated  in the limit $|x-y|\to\infty$ as
\eqn{
\Delta_{xy} & =&\frac{1}{\langle\hat{L}^{\dagger}\hat{L}\rangle^{2}}\left(\langle\sum_{z,w}\hat{l}_{z}^{\dagger}\hat{l}_{w}\rangle\langle\sum_{z,w}\hat{l}_{z}^{\dagger}\hat{h}_{x}\hat{h}_{y}\hat{l}_{w}\rangle-\langle\sum_{z,w}\hat{l}_{z}^{\dagger}\hat{h}_{x}\hat{l}_{w}\rangle\langle\sum_{z,w}\hat{l}_{z}^{\dagger}\hat{h}_{y}\hat{l}_{w}\rangle\right)\label{Dxy2}\\
 & \simeq&\frac{1}{|\langle\hat{L}\rangle|^{4}}\left[(\langle\hat{l}_{x}^\dagger\rangle\langle\hat{h}_x\rangle-\langle\hat{l}_{x}^{\dagger}\hat{h}_{x}\rangle)(\langle\hat{l}_y^\dagger\rangle\langle\hat{h}_y\rangle-\langle\hat{l}_{y}^{\dagger}\hat{h}_{y}\rangle)\langle\hat{L}\rangle^2+{\rm c.c.}\right]\propto {\cal O}\left(\frac{1}{V^{2}}\right).
}
We thus conclude that the standard deviation of  energy is subextensive:
\eqn{
\Delta E=\sqrt{\sum_{x}\Delta_{xx}+\sum_{x\neq y}\Delta_{xy}}\simeq\sqrt{\sum_{x}\Delta_{xx}}\propto {\cal O}(\sqrt{V}).
}

We next consider the other case in which the expectation value of $\hat{L}/V$ vanishes in the thermodynamic limit. To be specific, we impose the following condition
\eqn{\label{scaling}
\langle\hat{L}\rangle=\langle\sum_{z}\hat{l}_{z}\rangle\propto {o}(\sqrt{V}).
}
For instance, in the numerical example presented in the main text, an expectation value of $\hat{L}=\sum_{m}(-1)^m\hat{n}_m$ with respect to arbitrary energy eigenstate is exactly zero, and thus the condition~\eqref{scaling} is satisfied. Using this condition, we can rewrite $\langle\hat{L}^\dagger\hat{L}\rangle$ as
\eqn{
\langle\hat{L}^{\dagger}\hat{L}\rangle=\langle\sum_{z,w}\hat{l}_{z}^{\dagger}\hat{l}_{w}\rangle\simeq\sum_{z,w}(\langle\hat{l}_{z}^{\dagger}\hat{l}_{w}\rangle-\langle\hat{l}_{z}^{\dagger}\rangle\langle\hat{l}_{w}\rangle)\simeq\sum_{z}\left(\langle\hat{l}_{z}^{\dagger}\hat{l}_{z}\rangle-\langle\hat{l}_{z}^{\dagger}\rangle\langle\hat{l}_{z}\rangle\right)\propto {\cal O}(V).
}
Here, in the first approximate equality we add the $o(V)$ contribution in Eq.~\eqref{scaling}, and in the second approximate equality we use the CDP of $\hat{\rho}$ and the scaling relation~\eqref{scaling}.
The leading contribution in $\Delta_{xy}$ of Eq.~\eqref{Dxy2} is obtained as
\eqn{
\Delta_{xy}\simeq\frac{\langle\hat{l}_{x}^{\dagger}\hat{h}_{x}\rangle\langle\hat{h}_{y}\hat{l}_{y}\rangle+{\rm c.c.}}{\sum_{z}\left(\langle\hat{l}_{z}^{\dagger}\hat{l}_{z}\rangle-\langle\hat{l}_{z}^{\dagger}\rangle\langle\hat{l}_{z}\rangle\right)}\propto {\cal O}\left(\frac{1}{V}\right),
}
again leading to a subextensive energy fluctuation:
\eqn{
\Delta E=\sqrt{\sum_{x}\Delta_{xx}+\sum_{x\neq y}\Delta_{xy}}\propto {\cal O}(\sqrt{V}).
}

\subsection{Numerical results for a local continuous measurement}

\begin{figure}[t]
\includegraphics[width=80mm]{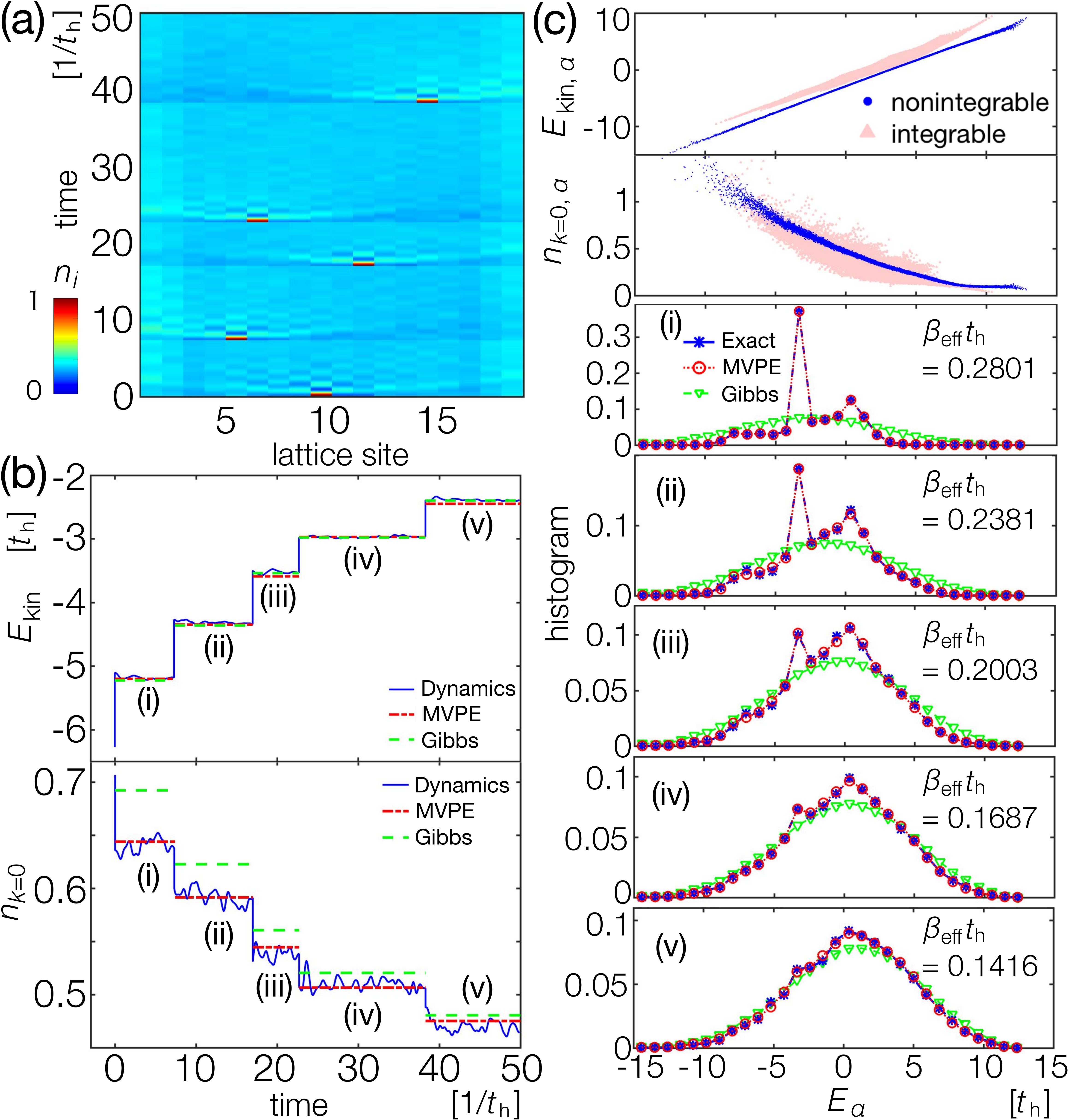} 
\caption{\label{figlocal}
(a) Spatiotemporal dynamics of the particle-number density in a typical quantum trajectory under the local measurement with $\hat{L}_l=\hat{n}_l$. Every time an atom is detected, a high-density region emerges which subsequently diffuses away toward an equilibrium value.  (b) The corresponding dynamics of the kinetic energy $\hat{K}$ and that of the occupation $\hat{n}_{k=0}$ at zero momentum. Superimposed are the predictions from the MVPE $\hat{\rho}_{\cal M}$ conditioned on a sequence of quantum jumps that have been observed until time $t$ (red dashed lines) and the Gibbs ensemble $\hat{\rho}_{\beta_{\rm eff}^{\cal M}}$ (green dashed lines). (c) Diagonal values of the observables in the energy basis (top two panels) and the energy distributions after each jump (the other panels).
We set $\gamma=0.02$ and $t_{\rm h}\!=\!U\!=\!t_{\rm h}'\!=\!U'\!=\!1$ while $t_{\rm h}\!=\!U\!=\!1$ and $t_{\rm h}'\!=\!U'\!=\!0$ for the integrable case in (c).
}
\end{figure}

We here present our numerical results on the many-body dynamics under the local measurement.
We consider the hard-core boson model discussed in the main text, which is nonintegrable and has been numerically confirmed to satisfy the ETH \cite{RM09,KH14,LDA16,YT18}. We set the system size and the total number of particles as $L_s\!=\!18$ and $N\!=\!6$. As a measurement process, we consider a local one $\hat{L}_{l}\!=\!\hat{n}_{l}$, where a quantum jump is labeled by a lattice site $l$. This measurement process can be implemented by the site-resolved position measurement of ultracold atoms via light scattering, as realized in quantum gas microscopy (microscopic derivations based on atom-photon interactions can be found in Refs.~\cite{PH10,YA15}).

Figure~\ref{figlocal} shows a typical trajectory dynamics under the local measurement. After each detection, measurement backaction localizes an atom on the detected lattice site, which subsequently spreads over and quickly relaxes toward an equilibrium density (Fig.~\ref{figlocal}a). Figure~\ref{figlocal}b shows the corresponding dynamics of the kinetic energy $\langle\hat{K}\rangle$ (top) and the occupation $\langle\hat{n}_{k=0}\rangle$ at zero momentum (bottom), and compares them with the predictions from the MVPE $\hat{\rho}_{\cal M}$ (red chain) and the Gibbs ensemble $\hat{\rho}_{\beta_{\rm eff}^{\cal M}}$ (green dashed). For each time interval, the dynamical values agree with the MVPE predictions within time-dependent fluctuations. 
Note that the MVPE is time independent by definition; the plotted values are the ones obtained from the MVPE conditioned on a sequence of quantum jumps that have been observed until time $t$. 

To gain further insights, Fig.~\ref{figlocal}c  plots the diagonal matrix elements of each observable in the energy basis (top two panels) and energy distributions after every quantum jump (the other panels). Small eigenstate-to-eigenstate fluctuations in observables and the remarkable agreement in the energy distribution demonstrate the validity of the MVPE description in Fig.~\ref{figlocal}b.
It is remarkable that only a few jumps are sufficient to smear out the initial memory of a single eigenstate after which the distribution is almost indistinguishable from that of the corresponding Gibbs ensemble, thus validating the relation~(9) in the main text.  Small fluctuations after the first jump (Fig.~\ref{figlocal}b) indicate that even a single quantum jump generates a sufficiently large effective dimension to make the system equilibrate, which can be understood from the substantial delocalization of an energy eigenstate in the Fock basis \cite{NC12}. To avoid a finite-size effect (which leads to a  discrepancy of the Gibbs ensemble from $\langle\hat{n}_{k=0}\rangle(t)$), it may be advantageous to use  $\hat{\rho}_{\cal M}$ rather than $\hat{\rho}_{\beta_{\rm eft}^{\cal M}}$ for small systems that can be prepared in experiments.

\subsection{Finite-size scaling analysis}
\begin{figure}[t]
\includegraphics[width=90mm]{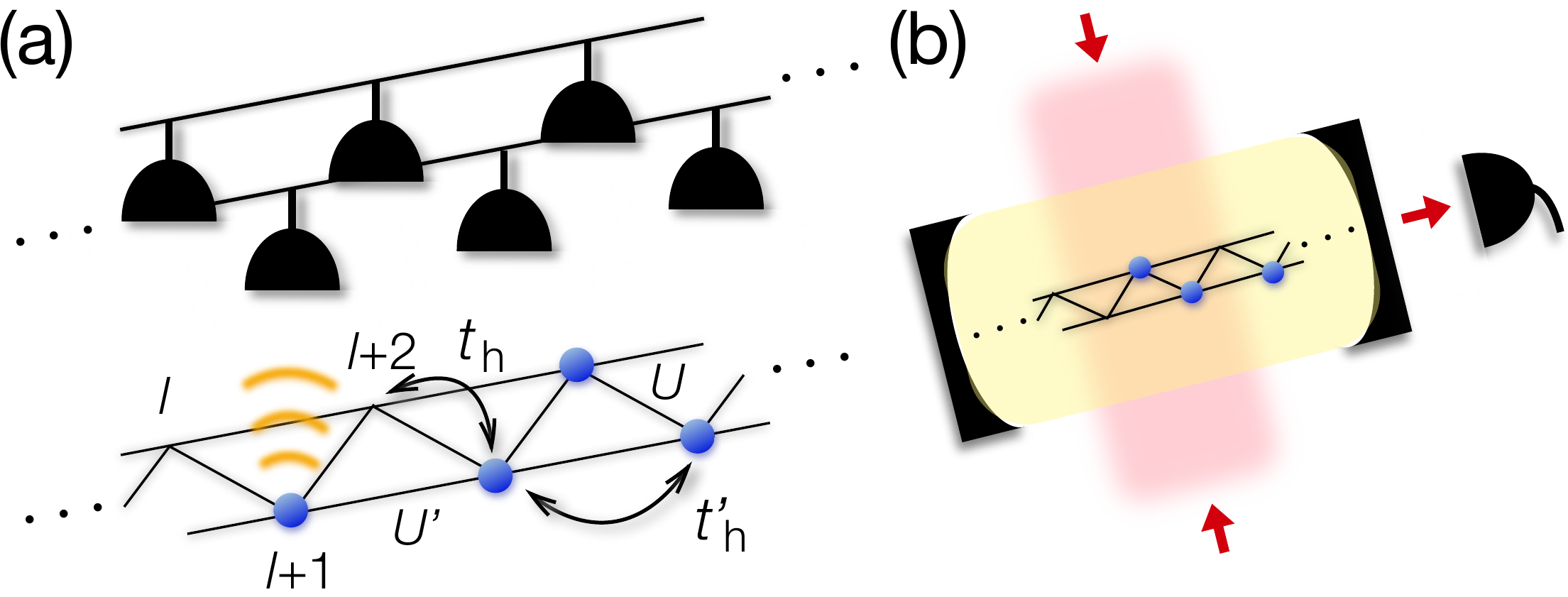} 
\caption{\label{figs1}
Quantum many-body systems under (a) local and (b) global continuous measurements. (a) Hard-core bosons on a lattice are subject to site-resolved position measurement via light scattering. (b) The cavity photon field is coupled to a collective mode of atoms and  photons emanating from the cavity are continuously monitored.
}
\end{figure}

In this section, we present the results of the finite-size scaling analysis for testing the precision of the predictions from the matrix-vector product ensemble (MVPE) and the corresponding Gibbs ensemble with an effective temperature.
Here and henceforth, we apply our theory to the specific model of one-dimensional hard-core bosons discussed in the main text. The system Hamiltonian $\hat{H}=\hat{K}+\hat{U}$ consists of the kinetic energy $\hat{K}$ and the interaction term $\hat{U}$, which involve the nearest- and next-nearest-neighbor hopping and an interaction (see Fig.~\ref{figs1}):
\eqn{
\hat{K}&=&-\sum_{l}(t_{{\rm h}}\hat{b}_{l}^{\dagger}\hat{b}_{l+1}\!+\!t'_{{\rm h}}\hat{b}_{l}^{\dagger}\hat{b}_{l+2}\!+\!{\rm H.c.}),\label{h1}\\ 
\hat{U}&=&\sum_{l}(U\hat{n}_{l}\hat{n}_{l+1}\!+\!U'\hat{n}_{l}\hat{n}_{l+2}),\label{h2}
}
where $\hat{b}_l$ ($\hat{b}_{l}^\dagger$) is the annihiliation (creation) operator of a hard-core boson on site $l$ and $\hat{n}_{l}\!=\!\hat{b}_{l}^\dagger\hat{b}_{l}$. We use the open boundary conditions. To be specific, when we consider a nonintegrable case, we set $t_{\rm h}=t_{\rm h}'=U=U'=1$ for which the system is known to satisfy the ETH \cite{RM09,KH14,LDA16,YT18}. We consider  local or global continuous measurements performed on this many-body system. The local measurement is associated with a jump operator $\hat{L}_{l}=\hat{n}_l$ acting on the lattice site $l$, which physically corresponds to a site-resolved position measurement of atoms in an optical lattice as realized in quantum gas microscopy (see Fig.~\ref{figs1}(a)) \cite{BWS09,SJF10,PH10,YA15}. The global measurement is associated with a jump operator $\hat{L}=\sum_{l}(-1)^l\hat{n}_l$ acting on an entire region of the system, which can be realized by continuously monitoring photons leaking out of a cavity coupled to a certain collective atomic mode (see Fig.~\ref{figs1}(b)) \cite{MG16}.

\begin{figure}[t]
\includegraphics[width=180mm]{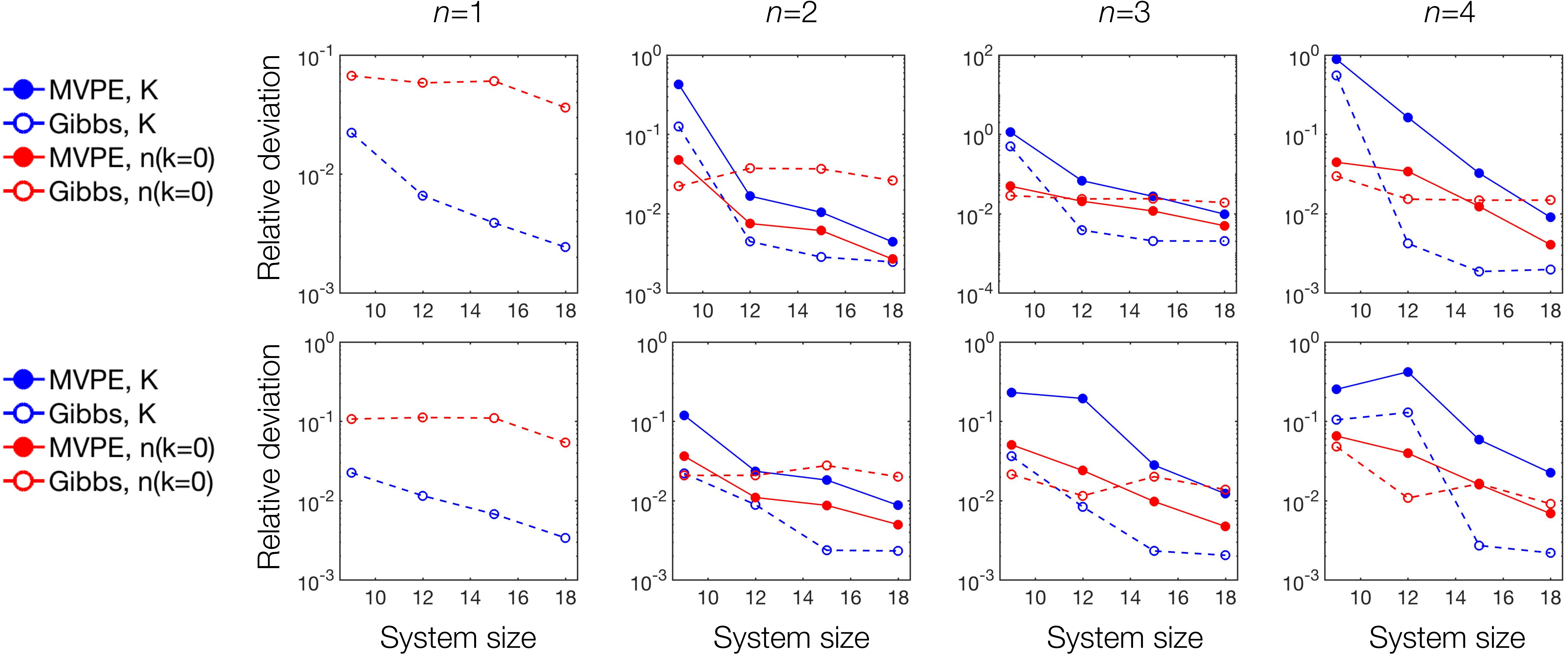} 
\caption{\label{figfs}
Finite-size scaling analyses of the relative deviation of the predictions of the MVPE and the corresponding Gibbs ensemble from the time-averaged value. The top (bottom) panels show the relative errors of the observables for each ensemble from their time-averaged values in the trajectory dynamics after the $n$-th quantum jump with the local (global) measurement. The relative deviations of the MVPE values from the time-averaged ones are plotted as the blue (red) filled circles for the kinetic energy $\hat{K}$ (the occupation number at  zero momentum $\hat{n}_{k=0}$). In the same way, the deviations of the corresponding Gibbs ensemble are plotted as the open circles. We note that the deviations of the MVPE predictions after the first jump $n=1$ are negligibly small and not shown in the plots. We set $\gamma=0.02$ and $t_{\rm h}=t_{\rm h}'=U=U'=1$.
}
\end{figure}

We numerically simulate quantum trajectory dynamics starting from the energy eigenstate $|E_0\rangle$ corresponding to the initial temperature $T_0=3t_{\rm h}$ as explained in the main text. To perform the finite-size scaling analysis, we calculate the relative deviation of the predictions of the MVPE or the corresponding Gibbs ensemble from the time-averaged value:
\eqn{
r_{\hat{\rho}}=\frac{\left|\overline{\hat{O}(t)}-\langle\hat{O}\rangle_{\hat{\rho}}\right|}{\left|\overline{\hat{O}(t)}+\langle\hat{O}\rangle_{\hat{\rho}}\right|},
}
where $\overline{\hat{O}(t)}$ denotes the time-averaged value of an observable $\hat{O}$ over the trajectory dynamics during the time interval involving $t$  between quantum jumps, $\langle\cdot\rangle_{\hat{\rho}}={\rm Tr}[\cdot\hat{\rho}]$ with $\hat{\rho}$  chosen to be either the MVPE $\hat{\rho}_{\cal M}$ or the Gibbs ensemble $\hat{\rho}_{\beta_{\rm eff}^{\cal M}}$ with an effective temperature $\beta_{\rm eff}^{\cal M}$. As an observable $\hat{O}$, we use the kinetic energy $\hat{K}$ or the occupation number $\hat{n}_{k=0}$ at  zero momentum. We set the filling $N/L=1/3$ with $N$ and $L$ being the total number of atoms and the system size, and vary $N$ from $3$ to $6$. The results are presented in Fig.~\ref{figfs}. The top (bottom) panels show the relative errors $r_{\hat{\rho}}$ for each time interval after the $n$-th jump event in the trajectory dynamics with the local (global) measurement processes. These finite-size scaling analyses indicate that the relative errors of the MVPE predictions (filled circles) converge almost exponentially to zero in the thermodynamic limit. We find that the convergence of the corresponding Gibbs ensemble predictions (open circles) is slower than that of the MVPE. This fact can be attributed to a combination of  broad energy distributions of finite-size systems and large fluctuations in diagonal elements of observables (see e.g., Fig.~\ref{figlocal}(c)). It is worthwhile to mention that a similar slow convergence of the observable $\hat{n}_{k=0}$ to the equilibrium value due to finite-size effects has also been  found in the time-dependent density-matrix renormalization-group calculations of the Bose-Hubbard model with spontaneous emissions \cite{SJ14}. We note that the results presented in Fig.~2b in the main text are obtained by taking the averages over the values for different jumps $n=2,3,4$ shown in Fig.~\ref{figfs}.

\subsection{Numerical analysis on integrable systems}

\begin{figure}[p]
\includegraphics[width=120mm]{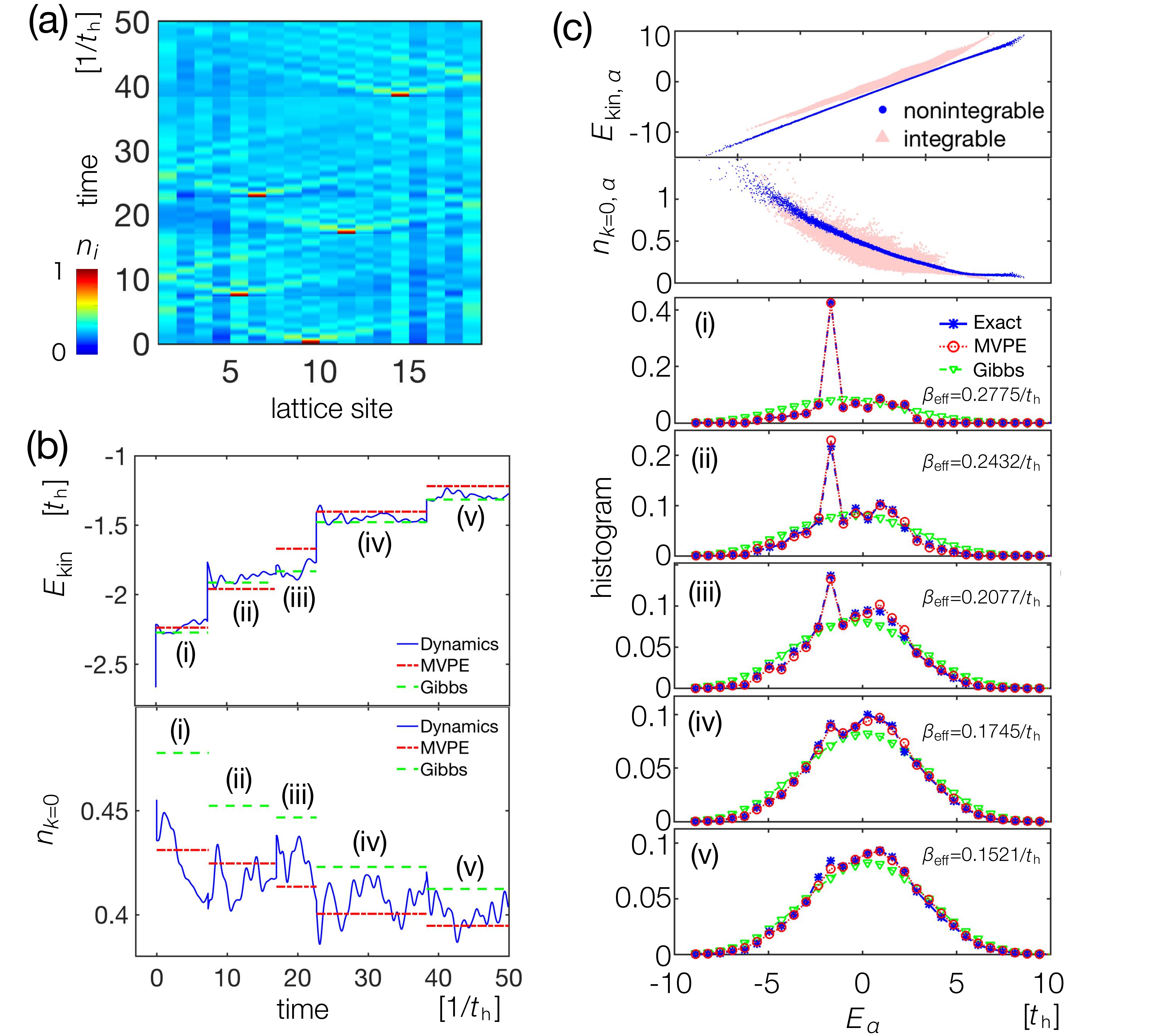} 
\caption{\label{figint}
Numerical results on the trajectory dynamics for which the system Hamiltonian is integrable and the measurement is local. (a) Spatiotemporal dynamics of the atom-number distribution at each lattice site and (b) the corresponding dynamics of the kinetic energy $\hat{K}$ (top panel) and the occupation number at  zero momentum $\hat{n}_{k=0}$ (bottom panel). (c) Top two panels show the diagonal values of each observable in the energy basis for the integrable case (red triangle) and the nonintegrable case (blue circle). Other panels show the changes of energy distributions after each jump. The system size is $L_s=18$ and the total number of bosons is $N=6$. We set  $\gamma=0.02$ and $t_{\rm h}=U=1$ and $t_{\rm h}'=U'=0$ for the integrable system Hamiltonian, and set $t_{\rm h}=t_{\rm h}'=U=U'=1$ to plot nonintegrable results in the top two panels of (c).
}
\end{figure}

An isolated integrable many-body system often fails to thermalize because the Gibbs ensemble is not sufficient to fix distributions of an extensive number of local conserved quantities.
Here we present numerical results of trajectory dynamics with an integrable  system Hamiltonian. To be specific, we consider a local measurement process $\hat{L}_l=\hat{n}_l$ and choose the parameters as $t_{\rm h}=U=1$ and $t_{\rm h}'=U'=0$. For the sake of comparison, in Fig.~\ref{figint} we present the results for the trajectory dynamics with the same occurrence times and types of quantum jumps as realized in the nonintegrable results presented in Fig.~\ref{figlocal}. The initial state is again chosen to be the energy eigenstate $|E_0\rangle$ of the integrable many-body Hamiltonian having the energy corresponding to the temperature $T_0=3t_{\rm h}$.

Figure~\ref{figint}(a) shows the spatiotemporal dynamics of the atom number at each lattice site. Measurement backaction localizes an atom at the site of detection and the density waves propagate ballistically through the system. The induced density fluctuations are significantly larger than those found in the corresponding nonintegrable results, and the relaxation to the equilibrium profile seems to be not reached during each time interval between quantum jumps in the integrable case. Also, the ballistic propagations of density waves are reflected back at the boundaries and can disturb the density; the finite-size effects can be more significant in the integrable case than the corresponding nonintegrable one. It merits further study to identify an equilibration time scale in an integrable many-body system under continuous measurement. 
Figure~\ref{figint}(b) shows the corresponding dynamics of the kinetic energy $\hat{K}$ and the occupation number $\hat{n}_{k=0}$ at  zero momentum. Relatively small (large) time-dependent fluctuations in the kinetic energy (the zero-momentum occupation) can be attributed to the small (large) fluctuations  of its diagonal values in the energy basis (see the top two panels in Fig.~\ref{figint}(c)). Other panels in Fig.~\ref{figint}(c) show the corresponding changes of energy distributions after each quantum jump. In a similar manner as in the nonintegrable results presented in Fig.~\ref{figlocal},  a few jumps are enough to smear out the initial memory as a single energy eigenstate. This fact implies that the biased weight on a possible nonthermal eigenstate admitted by the weak variant of the ETH \cite{BG10,TM16} will disappear after observing a few number of quantum jumps. The jump operator $\hat{L}_l$ acts as a weak-integrability breaking (nonunitary) perturbation and should eventually make the system thermalize. 

The thermalization behavior can be also inferred from the eventual agreement between the time-dependent values of the observables and the predictions from the Gibbs ensemble after several jumps (see Fig.~\ref{figint}(b)). Nevertheless, it is still evident that a largely biased weight on an initial (possibly nonthermal) state can survive when a number of jumps is small (see e.g., the panels (i) and (ii) in Fig.~\ref{figint}(c)), and thus the generalized Gibbs ensemble can be a suitable description for such a case. To make concrete statements, we need a larger system size and more detailed analyses with physically plausible initial conditions. It remains an interesting open question to examine to what extent the initial memory of a possible nonthermal state can be kept under an integrability-breaking  continuous measurement process.   

\end{document}